\documentclass[sn-chicago]{sn-jnl}% Chicago-based Humanities Reference Style

\usepackage{graphicx}%
\usepackage{multirow}%
\usepackage{amsmath,amssymb,amsfonts}%
\usepackage{amsthm}%
\usepackage{mathrsfs}%
\usepackage[title]{appendix}%
\usepackage{xcolor}%
\usepackage{textcomp}%
\usepackage{manyfoot}%
\usepackage{booktabs}%
\usepackage{algorithm}%
\usepackage{algorithmicx}%
\usepackage{algpseudocode}%
\usepackage{listings}%
\usepackage{multirow}
\usepackage{amsthm}

\newcommand{\vect}[1]{\boldsymbol{#1}}

\theoremstyle{thmstyleone}%
\newtheorem{theorem}{Theorem}%  meant for continuous numbers
\theoremstyle{thmstyletwo}%

\theoremstyle{thmstylethree}%
\newtheorem{definition}{Definition}%

\begin{document}

\title[The $\infty$-S test via quantile affine LASSO ]{The $\infty$-S test via quantile affine LASSO }


%% \author*[1]{\fnm{Sylvain} \sur{Sardy}}\email{Sylvain.Sardy@unige.ch}

%% \author[1]{\fnm{Ivan} \sur{Mizera}}\email{mizera@karlin.mff.cuni.cz}

%% \author[2]{\fnm{Xiaoyu} \sur{Ma}}\email{xyu.ma@outlook.com}
%% %\equalcont{These authors contributed equally to this work.}

%% \author[3]{\fnm{Hugo} \sur{Gaible}}\email{hugo.gaible@ens-paris-saclay.fr}

%\affil*[1]{\orgdiv{Department of Mathematics, Switzerland}, \orgname{University of Geneva}, \orgaddress{\street{Street}, \city{City}, \postcode{100190}, \state{State}, \country{Switzerland}}}
%
%\affil[2]{\orgdiv{National University of Defense and Technology}, \orgname{Organization}, \orgaddress{\street{Street}, \city{City}, \postcode{10587}, \state{State}, \country{China}}}
%
%\affil[3]{\orgdiv{Ecole Normale Sup\'erieure Paris Saclay}, \orgname{Organization}, \orgaddress{\street{Street}, \city{City}, \postcode{610101}, \state{State}, \country{France}}}

\author[1]{\fnm{Sylvain} \sur{Sardy}}\email{Sylvain.Sardy@unige.ch}

\author[2]{\fnm{Ivan} \sur{Mizera}}\email{mizera@karlin.mff.cuni.cz}

\author*[3]{\fnm{Xiaoyu} \sur{Ma}}\email{xyu.ma@outlook.com}

\author[4]{\fnm{Hugo} \sur{Gaible}}\email{hugo.gaible@ens-paris-saclay.fr}

%%% \affil[1]{\orgdiv{Department of Mathematics, Switzerland}}
\affil[1]{\orgdiv{University of Geneva, Section of Mathematics, Switzerland}}

\affil[2]{\orgdiv{ Charles University Prague, Faculty Mathematics and Physics, Czechia}}

\affil[3]{\orgdiv{National University of Defense and Technology, College of Science, China}}

\affil[4]{\orgdiv{\'Ecole Normale Sup\'erieure Paris-Saclay, Department of Mathematics, France}}


\abstract{A novel test in the linear $\ell_1$ (LAD) and quantile
  regressions is proposed, based on the scores provided by the dual
  variables (signs) arising in the calculation of the (so-called)
  affine-lasso estimate---a Rao-type, Lagrange multiplier test using
  the thresholding, towards the null hypothesis of the test, function
  of the latter estimate.}




\maketitle

\section{Introduction}
We propose a new statistical test of a linear hypothesis in the linear
$\ell_1$ (LAD) and quantile
regression. In that context, the inference is considerably
nonparametric, as the null hypothesis in testing is as a rule
expressed via certain restriction on the median or pertinent quantile.
For instance, the well-known sign test, going back to John Arbuthnot
in 1710 \citep{Spr89,Con99}, tests the null hypothesis that the common
median of independent observations $Y_i$ is $\mu$; the test statistic
is then the number of positive signs among the signs of all $Y_i -
\mu$ (in practice, one would like to avoid zeros of $Y_i-\mu$, but
that is a minor complication that can be dealt with). Apart from the
independence assumption, the distributional specification is quite
loose; the $Y_i$'s even do not have to have the same distribution.
Thus, there is a lot of peace of mind in the application of the test,
as one has not to put too much faith in overly detailed assumptions.
In the traditional terminology, the sign test is dubbed
\emph{nonparametric}, meaning that it is guaranteed to maintain the
nominal level under fairly weak assumptions---and at the same time
exhibits reasonable power against numerous alternatives; of course,
here being nothing for free, the potential cost of such generality may
be the loss of efficiency compared to certain parametric tests derived
for exactly specified distributions---\emph{but only} rather in case
when the data really follow those.

It is well-known now that the use of the $\ell_1$ loss (cost) function
in regression predated the introduction of the $\ell_2$ one by Gauss
and Legendre by about half a century. The computation feasibility of
the latter, combined with the flexibility and plausibility of the
distributional assumption of normality, eclipsed the early 1760
invention of Boscovich for the centuries to come. However, the
computational breakthroughs in linear programming, and among all, the
possibility of broadening of the scope of the $\ell_1$ approach of the
LAD (``Least Absolute Deviations'') regression to quantile regression,
the study of conditional quantiles championed by \cite{KB78}, brought
the pendulum of the attention considerably back.

This gained an additional strong momentum when the $\ell_1$ loss
function emerged also in the penalized formulations of LASSO (Least
Absolute Shrinkage and Selection Operator, also known as Basis
Pursuit)%
%% \footnote{In Qu\'ebec, OSERMA, Op\'erateur de Selection Et
%% R\'etr\'ecissement de Moyen Absolu}
%
\citep{CDS99,Tibs:regr:1996}.
While the original formulations retained the $\ell_2$ loss in the
lack-of-fit term, to maintain a bond with the prevailing methodology,
an attractive synthesis arose in the quantile regression LASSO, its
particular median case known as LAD-LASSO \citep{CIS-215377}; the
latter framework carries a conceptual advantage that both lack-of-fit
term and penalty term feature the same form of the loss functions---so
that the LAD-LASSO can be written in an augmented $\ell_1$ regression
form.
%\citet{he2023},

All that said, the problem of inference in the $\ell_1$ regression
is indisputably appealing. The large-sample approach dates back to
\citet{kolmog31}, who already established the asymptotic normality of
the sample median for independent, identically distributed
observations; this was extended to the regression case by
\citet{LADasym91}, and refined by others. The main hurdle of the
implementation of this approach is that the asymptotic variance
depends on the reciprocal of the density (that is, ``sparsity'') at
the (potentially unknown) median of the distribution of the
observations (the distribution of the errors in the regression case).
The large-sample Wald-type test is then bound to estimate this unknown
nuisance parameter; while this is not an impossible task---various
strategies proposed in this respect are reviewed in the book of
\citet{Koenker}---it may still constitute a hurdle one would like to
bypass.

And indeed: the Rao's ``score'' approach to testing (also known as
Lagrange multiplier test), elaborated for inference in the
$\ell_1$/quantile regression by \citet{Roger93} via so-called
regression rank scores \citep{gutejure92}, the dual variables of the
primal convex optimization task defining the estimates, brought
far-reaching generalizations of the traditional one-sample and
two-sample rank tests (including, in particular, the sign test). While
it is a general wisdom that these methods avoid the estimation of
sparsity---as underlined explicitly by \citet{Koenker}%
%\footnote{page 92}
%
in the context of confidence intervals---\citet{he2023} mention
computational instabilities for large samples and various other
problems, and also the estimation of asymptotic variance which could
allegedly mar the rank test approach---quoting, however, rather the
more abstract account of \citet{gutejure92} instead of the more
focused \citet{Roger93} .

The problem here is pretty much in the eye of beholder, however.
Unlike the simple hit-and-miss paradigm of the sign-test, rank tests
allow for various flavors depending on what kind of rank scores are
adopted; the scores make Wilcoxon tests different from the van der
Waerden ones, for example. A human investigator may prefer one to
another for the one- and two-sample problems, and accordingly may opt
for the corresponding scores also in the $\ell_1$/quantile setting.
For instance, van der Waerden scores may be preferred by those who
believe that the reality is mostly normal (``Gaussian''), apart from
occasional erratic disturbances. Thus, once the flavor of the rank
test is chosen \emph{a priori}, the procedure ``does not require any
nuisance parameter depending on the error distribution to be
estimated'' \citep{Roger93}.%\footnote{page 308}.

However, when the rank testing is to be performed in an automatic way,
the rank scores may have to be elucidated in an ``adaptive way'',
using the existing theory of optimality of certain scores for certain
distributions (for example, as mentioned above, van der Waerden scores
work best for the normal distribution) to estimate the right scores
directly from the data. Ascending this next step on the ladder brings
a need of distributional estimation, potentially including that of
sparsity, back into the game.

The desire for automatic, human-free procedures may thus favor a
simple solution without a need of additional tuning choices; if such a
solution is available, and exhibits decent power compared to the
already existing options, it possesses a definite raison-d'\^etre, the
aspect we aim at demonstrating below. Section~\ref{sct:Stest} gives
the derivation of $\infty$-S test by defining the affine LAD LASSO,
elucidating its zero-thresholding function, and subsequently proposing
the sign-based test statistic that is asymptotically pivotal with
respect to the nuisance parameters. In Section~\ref{sct:opti}, we
discuss optimization issues related to LAD regression with linear
constraints and affine LAD LASSO; Section~\ref{sct:qStest} extends the
$\infty$-S test to testing a quantile of the distribution through the
definition of the quantile affine LASSO estimator. In
Section~\ref{sct:appli}, the $\infty$-S test is applied in four
settings: two simple settings to retrieve existing sign tests, a
quantile total variation sign test to test for a jump in the tail of
time series, and an empirical robustness analysis to non-Gaussianity
of the $\infty$-S test and the F-test to match the desired nominal
level of the test. The proofs are postponed to an Appendix.

\section{The $\infty$-S test} \label{sct:Stest}

Consider a linear model
\begin{equation}\label{eq:linearmodel}
\vect{y}= X\vect{\beta}+\vect{e},
\end{equation}
where $X$ is an $n\times p$ matrix
%and $\vect{y}\in{\mathbb R}^n$ is a response
and $\vect{\beta}$ is a $p\times 1$ vector of unknown parameters.
%$\vect{e}$ are errors i.i.d.~sampled from a Gaussian
%distribution ${\rm N}(0,\sigma^2)$.
Given an $m\times p$ matrix $A$ of rank $m$ and a vector
$\vect{b}\in{\mathbb R}^m$, we consider testing
\begin{equation}\label{eq:H0H1}
 	H_0:\ A \vect{\beta}=\vect{b} \quad \text{against}\quad H_1:\ A\vect{\beta}\neq\vect{b}.
\end{equation}
A new testing procedure, which we propose to call an $\infty$-S
test, is based on the least absolute deviation version of the affine
LASSO point estimator \citep{SARDY2022107507}, a regularized estimator
that can be defined in the constrained form as a solution of the
problem
\begin{equation}\label{eq:CONLL}
	\| \vect{y}- X\vect{\beta}\|_1 \to \min_{\vect{\beta}\in{\mathbb R}^{p}}!
        \qquad\text{subject to}\qquad
        \| A\vect{\beta}-\vect{b}\|_1 \leq \Lambda.
\end{equation}
The ``Lagrangian'' form of the definition used by
\citet{SARDY2022107507} proclaims the same estimator a
solution to
\begin{equation}\label{eq:LADLASSO}
	\| \vect{y}- X\vect{\beta}\|_1 + \lambda \|
        A\vect{\beta}-\vect{b}\|_1 \to \min_{\vect{\beta}\in{\mathbb R}^{p}}! 
\end{equation}
for a suitable Lagrange multiplier $\lambda$ pertaining to the
constraint in \eqref{eq:CONLL}; the larger $\lambda$, the more this
estimator thresholds towards zero the entries of
$A\vect{\beta}-\vect{b}$ corresponding to the null hypothesis $H_0$ in
\eqref{eq:H0H1}. That is, there is an interval $[\lambda_0, \infty)$
  for which the solution to~\eqref{eq:LADLASSO} satisfies the
  condition of $H_0$ for any $\lambda\in[\lambda_0, \infty)$. The
    smallest such $\lambda$ is a function of the data and has a closed
    form expression which we call the \emph{zero-thresholding
      function}.

 \smallskip

 \begin{theorem} \label{thm:ztf}
	%Let $\vect{\|A\|_1^{\rm col}}\in{\mathbb R}^p$ 
	%Let $\vect{a}\in{\mathbb R}^p$ be the $\ell_1$ norms of the $p$ columns of $A$ and
	%let ${\cal J}_A=\{ j\in\{1,\ldots,p\}: a_j \neq 0\}$. %, and let $X_{{\cal J}_A}$ be the columns of $X$ which indexes are in ${\cal J}_A$.
	Let 
	\begin{equation}\label{eq:betaH0}
	  \hat{\vect{\beta}}_{H_0} \in\arg
          \min_{ \vect{\beta}\in{\mathbb R}^{p}}
          \| \vect{y}-X \vect{\beta} \|_1 \quad
           \text{\rm subject to} \quad A \vect{\beta}=\vect{b}.
	\end{equation}
	%and $\vect{r}=X {\hat{\vect{\beta}}}_{H_0}-\vect{y}$ be the corresponding residuals.
	%Call ${\cal I}_0=\{ i\in\{1,\ldots,n\}: r_i=0\}$,
	%$(X^{\rm T})_{{\cal I}_0}$ the column of $X^{\rm T}$ which indexes are in ${\cal I}_0$,
	%let $\vect{x}_{{\cal I}_0}\in{\mathbb R}^{p}$ be the $\ell_1$ norms of the $p$ rows of $(X^{\rm T})_{{\cal I}_0}$
	%and let ${\vect u}:=(X^{\rm T})_{{\cal I}_0^{\rm c}}{\rm sign}({\vect r}_{{\cal I}_0^{\rm c}})$.
	The zero-thresholding function of the affine LAD-LASSO primal problem~\eqref{eq:LADLASSO}  is 
	\begin{equation}\label{eq:lambda0}
		\lambda_0(\vect{y}, X, A)=\|(AA^{\rm T})^{-1} A X^{\rm T}{\vect \omega} \|_\infty,
	\end{equation}
where ${\vect \omega} $ is the dual variable associated
to~\eqref{eq:LADLASSO}, the sign of
$X\hat{\vect{\beta}}_{H_0}-{\vect y}$ of the constrained LAD~\eqref{eq:betaH0}
whenever $X\hat{\vect{\beta}}_{H_0}-{\vect y} \ne 0$.
 \end{theorem}

\smallskip

As satisfying $H_0$ is equivalent to $\| A \vect{\beta}-\vect{b} \|=0$
for any norm, the result of the zero-thresholding
function~\eqref{eq:lambda0} is the %(scalar) Lagrange multiplier for
the constraint $\| A \vect{\beta}-\vect{b} \|=0$; with the
$\ell_1$-norm, the pertinent Lagrange multiplier is finite. Instead of
the $\ell_1$-norm, we could have used the $\ell_2$-norm corresponding
to affine LAD-group LASSO \citep{Yuan:Lin:mode:2006}; the dual of the
$\ell_2$-norm being the $\ell_2$-norm itself, the corresponding
zero-thresholding function would have been $\lambda_0(\vect{y}, X,
A)=\|(AA^{\rm T})^{-1} A X^{\rm T}{\vect \omega} \|_2$. More
generally, one can use the $\ell_q$-norm, which corresponding
zero-thresholding function is $\lambda_0(\vect{y}, X, A)=\|(AA^{\rm
  T})^{-1} A X^{\rm T}{\vect \omega} \|_{q/(q-1)}$.

The zero-thresholding function yields the test statistic of the
$\infty$-S test, which the following theorem shows is asymptotically
pivotal. The $\infty$-S test can be thus considered ``Lagrange
multiplier test''---that is, the Rao score test---
%% as the test statistic $S$
%% involves the score of the LAD loss under $H_0$,
akin to the rank tests developed by \citet{Roger93};
%%%% could be classified into the same category;
%---with two noticeable differences:
%% the test statistic $S$ uses the infinite norm instead of the weighted
%% $\ell_2$-norm, and its distribution does not require estimating the
%% sparsity function of \citet{Roger93}. The sparisty function is only
%% needed to make the statistic asymptotically $\chi^2$ under $H_0$, not
%% to make the statistic pivotal. Indeed, as Theorem~\ref{thm:pivot}
%% shows, the score is already asymptotically pivotal. There is an
%% important change of paradigm between the generalization of the sign
%% test by \citet{Roger93} and our new $\infty$-S test: their test has an
%% asymptotic listed distribution (e.g., $F$ or $\chi^2$), while our test
%% has an unlisted distribution. Nevertheless, with today's computer, one
%% can easily sample the distribution of the test statistic
%% $S=\lambda_0(\vect{Y},X,A)$ by Monte Carlo and estimate the test
%% critical value as the empirical $(1-\alpha)$-quantile of $S$. The
%% p-value is also estimated as the percentage of Monte Carlo simulated
%% random variables larger than $\lambda_0(\vect{y},X,A)$.
%
but, unlike the statistic $T_n$ of \citet[eq.~(2.9)]{Roger93}, the
$\infty$-S test does not require the integration of scores for dual
variables coming from different quantile regressions, and avoids also
the need to select the rank scores (and subsequent potential
estimation of underlying density characterizations to achieve the
optimality of those).

%In what follows, 
\smallskip

\begin{theorem} \label{thm:pivot}
	Let $\vect{Y}$ be the response random vector
under $H_0$. Assuming the LAD coefficient estimates are asymptotically
        Gaussian centered around the true coefficients, the
        test statistic $S=\lambda_0(\vect{Y},X,A)$ is asymptotically
        pivotal.
\end{theorem}

\smallskip

%
%In order to test $H_0$ we would like to have a test statistic with a distribution independent of nuisance parameters, that is, a statistic pivotal w.r.t. $\vect{\beta} \in {\rm ker(A)}$ and $\gamma>0$, the cale parameter. 
%

The asymptotic normality of the LAD coefficient estimates required by
Theorem~\ref{thm:pivot} is a well known fact---proved by
\citet{LADasym91}, and refined by others. We are therefore in
the\ position to define the $\infty$-S test now.

\smallskip

\begin{definition}[The $\infty$-S  test]
The $\infty$-S test function, to test~\eqref{eq:H0H1} in a linear
model ~\eqref{eq:linearmodel} at a prescribed level $\alpha\in(0,1)$,
is defined to be
	$$
	\phi({\vect{y}},X,A)=\left \{ 
	\begin{array}{rl}
		0 & \lambda_0(\vect{y},X, A)\leq c_{\alpha},\\
		1 & \text{\rm otherwise},
	\end{array}
	\right . ,
	$$
where $c_\alpha$ is selected so that the test has nominal level
$\alpha$.
\end{definition}

\smallskip

Based on Theorem~\ref{thm:pivot}, the level of the $\infty$-S test
matches asymptotically the nominal level~$\alpha$. The test itself is
being based on the dual variables for the $\ell_1$ minimization
problem; these variables attain values in $[-1,1]$ and amount to the
sign of residuals if those are not equal to zero. The insensitivity to
the magnitude of the residuals makes the test resistant to outlying
observations; nonetheless, in model-behaved situations, when the
error distribution is exactly normal, a more powerful $\infty$-rankS
test can be obtained by weighting the sign $\vect{\omega}$ of the
residuals by the rank of the absolute value of the respective
residuals.

In the applications, the sought p-values (or critical values
$c_{\alpha}$, if desired) are then easily obtained via Monte Carlo
sampling of $S=\lambda_0(\vect{Y},X,A)$, the technology which nowadays
provides satisfactory results even in the absence (so far) of a
knowledge of the distribution of test statistics in closed form
(related, say, to some traditional distributions).

%% Regarding the critical values $c_{\alpha}$, so far we have not been
%% able to elucidate some known, or related to known, relevant
%% distribution; however, nowadays it makes no problem to estimate the
%% critical value, or directly the p-value of the test, by Monte Carlo
%% sampling of $S=\lambda_0(\vect{Y},X,A)$.
%
%% where $S=\lambda_0(\vect{Y},X,A)$, as given in~\eqref{eq:lambda0},
%% and         
%% $$
%% F = \frac{(n-p)(A\hat{\vect{\beta}}
%%   - \vect{b})^{\rm T}[A(X^{\rm T}X)^{-1}A^{\rm T}]^{-1}(A\hat{\vect{\beta}} -
%%   \vect{b})}%
%% {m(\vect{Y}-X\hat{\vect{\beta}})^{\rm T}(\vect{Y}-X\hat{\vect{\beta}})},
%% $$
%        
%% 	where ${\vect y}$ is the data response.
%
%% let $c_\alpha=F_S^{-1}(1-\alpha)$, 
%% and $\vect{Y}$ is the response random vector for the linear model
%% under $H_0$ and the function $\lambda_0$ is given
%% in~\eqref{eq:lambda0}. The $\infty$-S test function is defined by
%% 	$$
%% 	\phi({\bf y},X,A)=\left \{ 
%% 	\begin{array}{rl}
%% 		0 & \lambda_0(\vect{y},X, A)\leq c_\alpha = F_S^{-1}(1-\alpha)  \\
%% 		1 & {\rm otherwise}
%% 	\end{array}
%% 	\right . ,
%% 	$$
%% 	where ${\vect y}$ is the data response.
%

From the practical point of view, we also recommend, in the spirit of
\citet{SardyISI08}, to perform preliminary rescaling of
$(A,\vect{b})$, in order to achieve homogeneous power---that is, the
power not favoring certain alternative hypotheses to the detriment of
the other ones. Indeed, the matrices $A$ and $X$ create
heteroskedasticity in the random vector $\vect{W}$ in the test
statistics $S=\|\vect{W} \|_\infty$ with $\vect{W} =|(AA^{\rm T})^{-1}
A X^{\rm T}{\vect \omega}|$ before taking the maximum of all entries.
Consequently, under $H_1$, the components of $\vect{W}$ with small
deviations will be hidden behind the components with large deviations.
To treat equally potential components of $A\vect{\beta}-\vect{b}$ not
being equal to zero, we propose the following.

\smallskip

\begin{definition}[Homopower rescaling]\label{def:homo}
Consider the $\infty$-S test with test statistic $S=\|\vect{W}
\|_\infty$, where $\vect{W} =|(AA^{\rm T})^{-1} A X^{\rm T}{\vect
  \omega}|$ and ${\vect \omega}$ are the sign of the residuals of the
constrained LAD under $H_0$. Let $F_{W_k}$ be the marginal cumulative
distribution function of the $k$th component of $\vect{W}$ for
$k=1,\ldots,m$. Letting $D_{(A,X,{\bf b},\alpha)}$ be the diagonal
matrix with $k$th diagonal element equal to
$d_k:=F_{W_k}^{-1}(1-\alpha)$, the homopower rescaling rescales $A$ to
$D_{(X,A,{\bf b},\alpha)}A$ and $\vect{b}$ to $D_{(X,A,{\bf
    b},\alpha)}\vect{b}$ before applying the test.
\end{definition}

\section{Optimization aspects} \label{sct:opti}

The objective function of the LAD optimization task
\begin{equation} \label{eq:LADopti}
 	%\hat{\vect{\beta}} \in\arg 
 	 \| \vect{y}-X\vect{\beta} \|_1 \to \min_{ \vect{\beta}\in{\mathbb R}^{p}}!
\end{equation}
is convex, and it is well known that the optimization can be
transformed into a linear program \citep{KB78}. The $\infty$-S test
requires solving the constrained LAD optimization~\eqref{eq:betaH0}
that can be transformed into an unconstrained LAD optimization of the
form~\eqref{eq:LADopti}.

\smallskip

\begin{theorem}\label{eq:constrainedLAD}
Solving~\eqref{eq:betaH0} is equivalent to solving~\eqref{eq:LADopti}
with $\vect{y}_{-A}=\vect{y}-X_1 A_1^{-1}\vect{b}$ and $X_{-A}=X_2
-X_1 A_1^{-1} A_2 $, where $A_1$ are $m$ linearly independent columns
of $A=:[A_1,A_2]$ and $X=:[X_1,X_2]$ is the corresponding
decomposition.
\end{theorem}

\smallskip

%Since $A$ has rank $m$, the contrained LAD optimization~\eqref{eq:betaH0} can be transformed into an uncontrained LAD optimization with a solution in ${\mathbb R}^{p-m}$. 
The proof is straightforward: choose $m$ linearly independent columns
of $A$, call them $A_1$ and the remaining ones $A_2$, so that $A=[A_1,
  A_2]$, permuting $\vect{\beta}=(\vect{\beta}_1, \vect{\beta}_2)$
accordingly. The linear constraints thus lead to
$\vect{\beta}_1=A_1^{-1}(\vect{b}-A_2 \vect{\beta}_2)$ and $
\vect{y}-X \vect{\beta}$ becomes $ \vect{y}_{-A}-X_{-A} \vect{\beta}_2
$ with $ \vect{y}_{-A}=\vect{y}-X_1 A_1^{-1}\vect{b}$ and $X_{-A}=X_2
-X_1 A_1^{-1} A_2 $.
%$\vect{y}-X_1 \vect{\beta}_1- X_2 \vect{\beta}_2=\vect{y}-X_1 A_1^{-1}\vect{b}+ X_1 A_1^{-1} A_2 \vect{\beta}_2)- X_2 \vect{\beta}_2=(\vect{y}-X_1 A_1^{-1}\vect{b})- (X_2 -X_1 A_1^{-1} A_2 ) \vect{\beta}_2$

The affine LAD-LASSO optimization~\eqref{eq:LADLASSO} can be also
transformed into an unpenalized LAD of the form~\eqref{sct:opti}, as
it is equivalent to solving
$$
\min_{{\vect \beta}\in{\mathbb R}^p}
\biggl\| \left ( \begin{array}{c} {\vect y} \\
  {\lambda {\vect b}} \end{array} \right )-
  \left ( \begin{array}{c} X
  \\ {\lambda A}  \end{array} \right )    {\vect \beta} \biggr\|_1.
$$
Another way to perform the constrained LAD
optimization~\eqref{eq:betaH0} is to solve the affine
\mbox{LAD-LASSO}~\eqref{eq:LADLASSO} for
$\lambda=\lambda_0(\vect{Y},X,A)$ of Theorem~\ref{thm:ztf} to
guarantee complete thresholding---that is, enforcing the linear
constraint; this is akin to the exact penalty method of \citet{PG89}.
The following theorem links the dual of the affine LAD-LASSO
optimization~\eqref{eq:LADLASSO} and the constrained LAD optimization.

\smallskip

\begin{theorem}\label{thm:dualomega}
	The dual variable $\vect{\omega}$ of Theorem~\ref{thm:ztf}
        corresponding to the primal affine LAD LASSO
        optimization~\eqref{eq:LADLASSO}  is the dual variable of  the
%        LAD optimization~\eqref{eq:LADopti} with  $\vect{y}_{-A}$ and $X_{-A}$ of Theorem~\ref{eq:constrainedLAD}.
        LAD optimization with $\vect{y}_{-A}$ and $X_{-A}$ of
        Theorem~\ref{eq:constrainedLAD}.
\end{theorem}

\smallskip

Summing up, the $\infty$-S test can be straightforwardly implemented
using the existing software---in particular, employing the {\tt R}
package {\tt quantreg} \citep{Rogerrq}.

\section{The $\infty$-S$^\tau$ test for quantile regression} \label{sct:qStest}

Given $\tau \in (0,1)$, let ${\vect q}^\tau$ be conditional
$\tau$-quantile vector ${\vect q}^\tau$ of a response vector ${\vect
  Y}\mid X$. The linear model assumes the form
\begin{equation}\label{eq:linearq}
	{\vect q}^\tau=X{\vect \beta},
\end{equation}
where $X$ is the regression matrix as before. To estimate the
parameters ${\vect{\beta}}$ from~${\vect y}$ arising from
the response random vector ${\vect Y}$, 
%stochastic framework ${\vect Y}\mid X$,
quantile regression~\citep{KB78} solves the optimization problem
\begin{equation}\label{eq:leastrho}
   \| \vect{y}-X\vect{\beta}\|_{\rho_\tau} \to \min_{\vect{\beta}\in{\mathbb R}^{p}}!
\end{equation}
The residuals $\vect{r}=\vect{y}- X\vect{\beta}$ are now subjected to
the objective function $\| \vect{r}\|_{\rho_\tau}=\sum_{i=1}^n
\rho_\tau(r_i)$, where
\begin{equation}\label{eq:tilted}
	%\rho_\tau(q)=(\tau-1) \sum_{y_i<q} (y_i-q)+\tau  \sum_{y_i\geq q} (y_i-q)
	\rho_\tau(r_i)=\left \{ \begin{array}{rl}(\tau-1)  r_i &\  r_i<0, \\
		\tau  r_i & \ r_i\geq 0, \end{array} \right .
\end{equation}
is the ``check function''---the tilted $\ell_1$, ``pinball'' loss.
Note that the special case of $\tau=1/2$ reduces to the LAD
regression, as $\| \vect{r}\|_{\rho_{1/2}}=\| {\vect r} \|_1/2$.
% for the property that the $\tau$ empirical quantile estimate $\hat q_\tau$ of $q_\tau=F_Y^{-1}$ is a solution to $\min_q \rho_\tau(q)$.

To test the null hypothesis against the alternative hypothesis, as
in~\eqref{eq:H0H1} but in the quantile-regression spirit, the quantile
$\infty$-S test does not refer to the median (well, unless
$\tau=1/2$), but rather to (any) $\tau$-quantile for some fixed $\tau
\in (0,1)$. We define accordingly the affine $\rho_\tau$-LASSO as a
solution to
\begin{equation}\label{eq:rhoLASSO}
  \rho_\tau( \vect{y}- X\vect{\beta}) + \lambda \| A\vect{\beta}-\vect{b}\|_1
  \to \min_{\vect{\beta}\in{\mathbb R}^{p}}!
\end{equation}
for fixed $\lambda>0$ and $\tau \in (0,1)$. The following theorem
gives the zero-thresholding function of this estimator.

\smallskip

\begin{theorem} \label{thm:ztfqr}
	%Let $\vect{\|A\|_1^{\rm col}}\in{\mathbb R}^p$ 
	%Let $\vect{a}\in{\mathbb R}^p$ be the $\ell_1$ norms of the $p$ columns of $A$ and
	%let ${\cal J}_A=\{ j\in\{1,\ldots,p\}: a_j \neq 0\}$. %, and let $X_{{\cal J}_A}$ be the columns of $X$ which indexes are in ${\cal J}_A$.
	For fixed $\tau\in(0,1)$, let 
	\begin{equation}\label{eq:betaH0rho}
	  \hat{\vect{\beta}}_{H_0}^\tau \in\arg
          \min_{ \vect{\beta}\in{\mathbb R}^{p}}
          \rho_\tau(\vect{y}-X \vect{\beta})
          \quad {\rm subject\  to} \quad A \vect{\beta}=\vect{b}.
	\end{equation}
	%and $\vect{r}=X {\hat{\vect{\beta}}}_{H_0}-\vect{y}$ be the corresponding residuals.
	%Call ${\cal I}_0=\{ i\in\{1,\ldots,n\}: r_i=0\}$,
	%$(X^{\rm T})_{{\cal I}_0}$ the column of $X^{\rm T}$ which indexes are in ${\cal I}_0$,
	%let $\vect{x}_{{\cal I}_0}\in{\mathbb R}^{p}$ be the $\ell_1$ norms of the $p$ rows of $(X^{\rm T})_{{\cal I}_0}$
	%and let ${\vect u}:=(X^{\rm T})_{{\cal I}_0^{\rm c}}{\rm sign}({\vect r}_{{\cal I}_0^{\rm c}})$.
	The zero-thresholding function of the affine $\rho_\tau$-LASSO primal problem~\eqref{eq:rhoLASSO}  is 
	\begin{equation}\label{eq:lambda0rho}
		\lambda_0^\tau(\vect{y}, X, A)=\|(AA^{\rm T})^{-1} A X^{\rm T}{\vect \omega}^\tau \|_\infty,
	\end{equation}
	where ${\vect \omega}^\tau $ is the dual variable associated
        to the linear programming solution of~\eqref{eq:rhoLASSO}.
%% which are the  weighted sign of
%% 	$X\hat{\vect{\beta}}_{H_0}^\tau-{\vect y}$ of the constrained minimum $\rho_\tau$~\eqref{eq:betaH0rho}.
\end{theorem}

\smallskip

The zero-thresholding function leads to the statistic of the
$\infty$-S$^\tau$ test; the following theorem shows that it is
asymptotically pivotal.

\smallskip

\begin{theorem} \label{thm:pivottau}
	Let $\vect{Y}$ be the response random vector
under $H_0$ and let $\tau\in(0,1)$. If the $\tau$-quantile regression
        parameter estimates are asymptotically normal with %Gaussian
        the limit distribution centered about their true values, the
        test statistic $S^\tau=\lambda_0^\tau(\vect{Y},X,A)$
        %, where $\vect{Y}$ is the response random vector under $H_0$
        is asymptotically pivotal.
\end{theorem}

The asymptotic normality %Gaussianity
of the quantile regression estimators, as established by \citet{KB78}
(their results including the $\ell_1$ results as a special case),
allows for the following definition of the $\infty$-S$^\tau$
test---which Theorem~\ref{thm:pivottau} shows has asymptotically the
nominal level~$\alpha$.

\smallskip

\begin{definition}[The $\infty$-S$^\tau$ test]
  Given $\tau\in(0,1)$, the $\infty$-S$^\tau$ test to
  test~\eqref{eq:H0H1} in the linear model~\eqref{eq:linearq} at a
  prescribed level $\alpha\in(0,1)$, is defined to be
	$$
	\phi({\vect y},X,A)=\left \{ 
	\begin{array}{rl}
		0 & \lambda_0^\tau(\vect{y},X, A)\leq c_\alpha^\tau \\
		1 & {\rm otherwise}
	\end{array}
	\right . ,
	$$
where $c^{\tau}_\alpha$ is selected so that the test has nominal level
$\alpha$.
\end{definition}

%% ??????????????????
%% Weighting the sign
%% $\vect{\omega}^\tau$ of the residuals by the ranks of the weighted
%% residuals $|\vect{r}| \cdot \rho_\tau(\vect{r})$ leads to the
%% $\infty$-rankS$^\tau$ test.

\section{Special cases and applications} \label{sct:appli}

Our simulations and applications rely on the R package {\tt quantreg}
\citep{Rogerrq}. We use the {\tt rq} function to calculate the dual
variables $\vect{\omega}$ involved in the zero-thresholding function.
In all our applications of the $\infty$-S test, the critical values
are estimated by Monte Carlo simulation with $10^4$ runs. The levels
and power curves of the tests are estimated based on $10^4$ simulated
data sets. We use the {\tt anova.rq} function to perform the $\chi^2$
sign test whenever it returns a p-value; otherwise the results of the
$\chi^2$ sign test cannot be reported.

\subsection{The paired $\infty$-S test}

Consider two paired measurements $\{(u_i,v_i)\}_{i=1,\ldots,n}$, a sample from the model
$$
U_i=\mu_i+\tilde \epsilon_i 	\quad {\rm and} \quad V_i=\mu_i+\delta+\breve \epsilon_i, \ i=1,\ldots,n,
$$
where the errors $\tilde \epsilon$ and $\breve \epsilon$ are assumed independent with a median equal to zero.
To test
$$
H_0: \delta=0 \quad {\rm against} \quad H_1: \delta \neq 0
$$ with the $\infty$-S test, we write the model as ${\vect
  y}=X\beta+{\vect \epsilon}$ with ${\vect y}={\vect v}-{\vect u}$,
$X={\vect 1}_n$, $\beta=\delta$, ${\vect \epsilon}={\tilde {\vect
    \epsilon}}-{\breve {\vect \epsilon}}$, $A=1$ and $b=0$. The
value of the test statistic is then
$$s=|\sum_{i=1}^n {\rm sign}({\vect
  y})|=|\sum_{i=1}^n {\rm sign}({\vect v}-{\vect u})|=|\sum_{i=1}^n
(1(v_i>u_i)-1(v_i<u_i))|.$$
Given that the statistic of the sign test statistic in this situation 
is
$$\tilde s=\sum_{i=1}^n 1(v_i>u_i)
\quad \text{and} \quad
\sum_{i=1}^n
1(v_i>u_i)+\sum_{i=1}^n 1(v_i<u_i)=n,
$$
we have that $s=|2\tilde s-n|$ and the
$\infty$-S test is equivalent to the sign test.

\subsection{The unpaired $\infty$-S test}

Consider two unpaired measurements $\{u_i\}_{i=1,\ldots,n}$ and $\{v_i\}_{i=1,\ldots,n}$, a sample from the model
$$
U_i=\mu+\tilde \epsilon_i 	\quad {\rm and} \quad V_i=\mu+\delta+\breve \epsilon_i, \ i=1,\ldots,n,
$$
where the errors $\tilde \epsilon$ and $\breve \epsilon$ are assumed independent with a median equal to zero.
To test
$$
H_0: \delta=0 \quad {\rm against} \quad H_1: \delta \neq 0
$$
with  the $\infty$-S test, one writes the model as ${\vect y}=X{\vect \beta}+{\vect \epsilon}$ with ${\vect y}^{\rm T}=({\vect u}^{\rm T},{\vect v}^{\rm T})$,
$$
X=\left ( \begin{array} {cc}  \begin{array} {c} {\vect 1}_n \\ {\vect 1}_n \end{array} &    \begin{array} {c} {\vect 0}_n \\ {\vect 1}_n \end{array}  \end{array} \right ),
$$ ${\vect \beta}=(\mu, \delta)^{\rm T}$, ${\vect \epsilon}^{\rm
  T}=({\tilde {\vect \epsilon}}^{\rm T},{\breve {\vect \epsilon}}^{\rm
  T})$, $A=(0,1)$ and $b=0$. One gets now
$\hat{\vect{\beta}}_{H_0}=(m_{\vect y},0)^{\rm T}$, and the
test-statistic value is
$$s=\max(0,|\sum_{i=1}^n 1(v_i>m_{\vect
  y})-\sum_{i=1}^n 1(v_i<m_{\vect y})|).
$$
We obtain the sign test for unpaired data, also known as the median
test \citep{brown1951median}.
\begin{figure}[!t]
	\centering
	\includegraphics[width=4.8in]{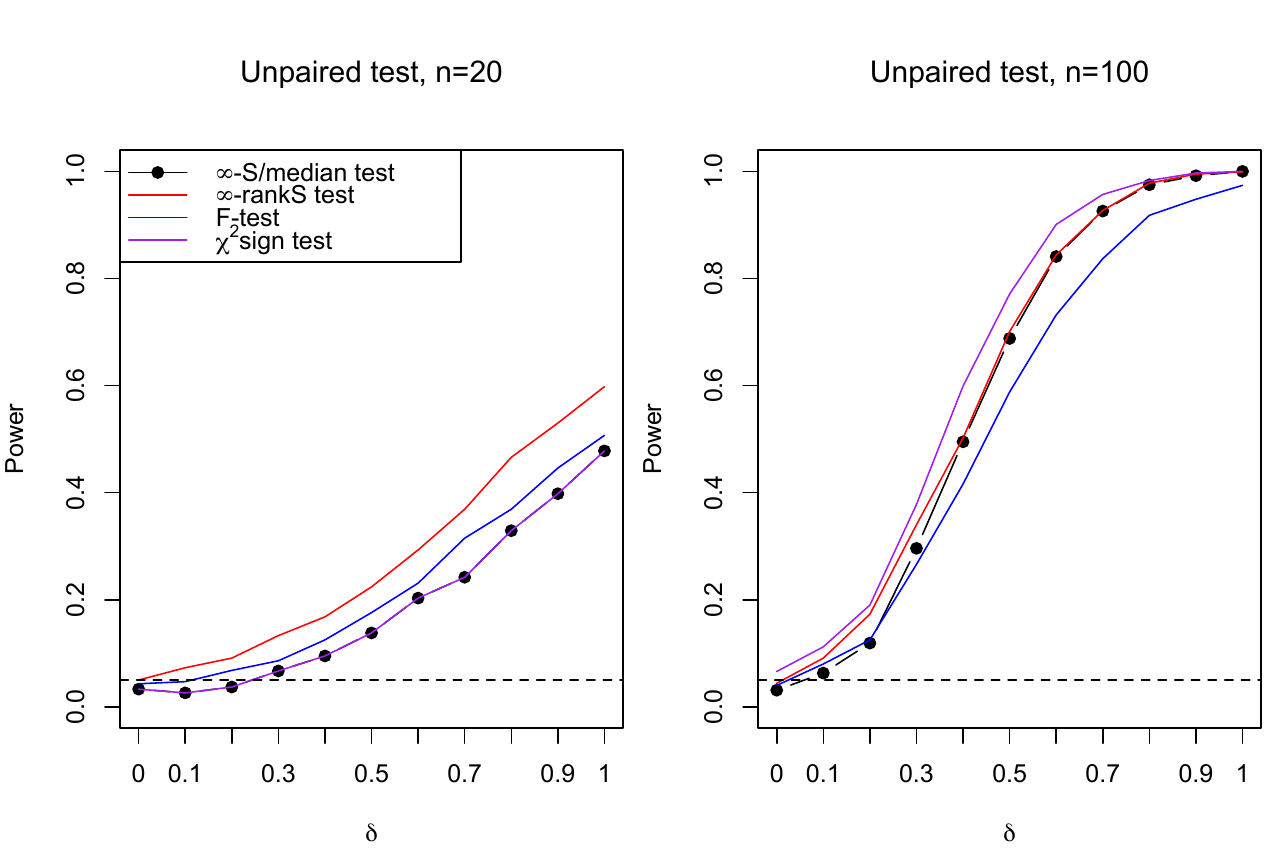}
	\caption{Power of the unpaired $\infty$-S$^\tau$ test (same as median test) and $\infty$-rankS$^\tau$ test for $\tau=0.5$, the F-test and asymptotically $\chi^2$ sign test  as a function of the shift $\delta\in[0,2]$ between the two populations. The horizontal dotted line is the nominal level $\alpha=0.05$. Sample size $n=20$ (left plot) and $n=100$ (right plot). }
	\label{fig:section52}
\end{figure}

We compare the level and the power of four tests ($\infty$-S or median test, $\infty$-rankS test, F-test and $\chi^2$ sign test) on two Monte Carlo simulations with $n=20$ and $n=100$ and with Student errors with 3 degrees of freedom.  Figure~\ref{fig:section52} shows that all tests satisfy the nominal level of $\alpha=0.05$ and the F-test looks quite robust to non-Gaussian errors although losing some power. The $\chi^2$ sign test seems to overshoot the level, even for $n$ large, however.
%Surprisingly, for $n=100$ the asymptotically $\chi^2$ sign-test overshoots the nominal level; we will see in other simulations in higher dimension that this test has difficulties performing well.

%%%%%%%%%%%%%%%%%%%%%%%%%%%%%
\subsection{Homopower rescaling}\label{sct:homopower}

\begin{figure}[!t]
	\centering
	\includegraphics[width=5.0in]{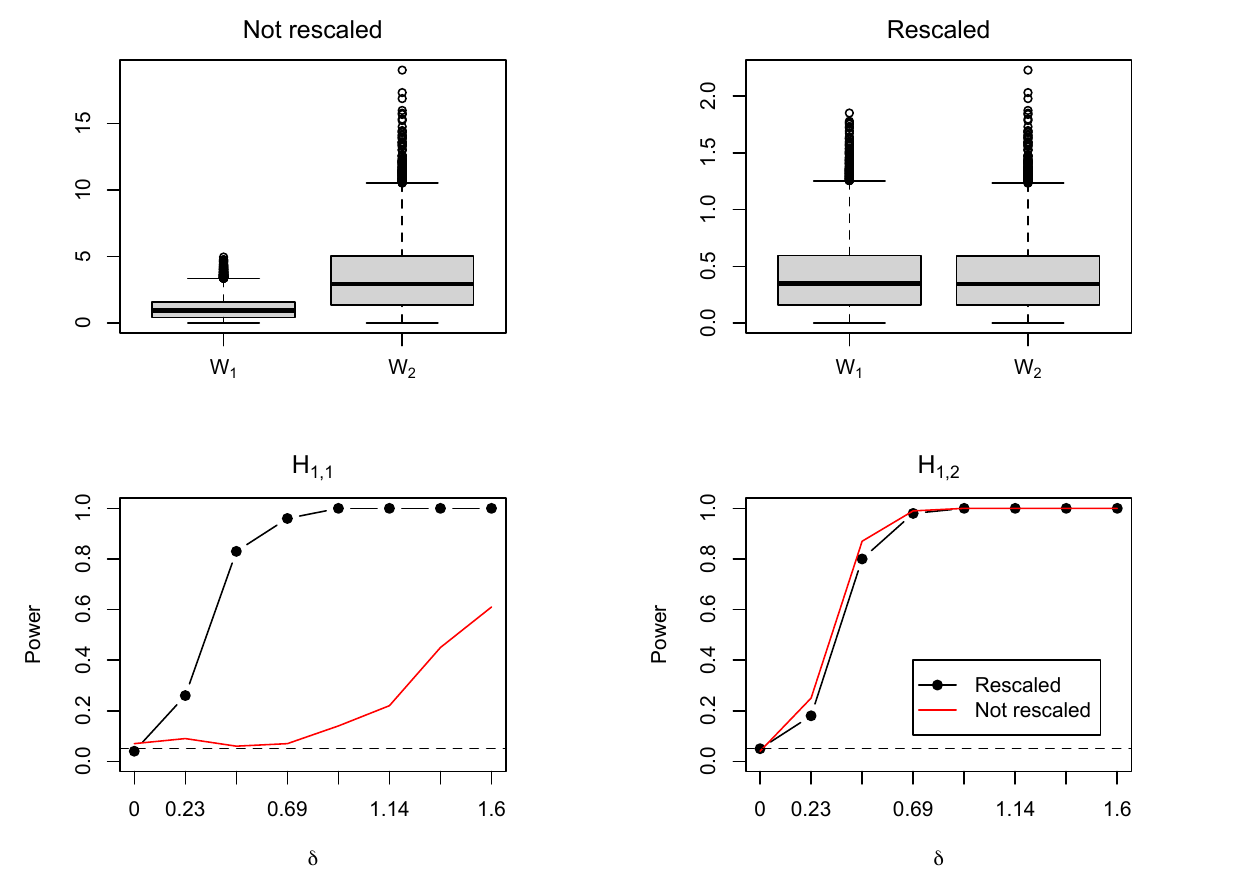}
	\caption{Homopower rescaling. Top: boxplots of Monte Carlo simulated  bivariate ($m=2$) random vector $W$ unrescaled (left) and rescaled (right).
		Bottom: power plots of the unrescaled (red) and rescaled (black)  $\infty$-S test under $H_{1,1}$ (left) and $H_{1,2}$ (right).}
	\label{fig:rescaleTF}
\end{figure}

We perform a small simulation study to outline the importance of
homopower rescaling (see Definition~\ref{def:homo}) of the two
components $A$ and $\vect{b}$ of the linear null
hypothesis~\eqref{eq:H0H1}. We set $n=100$, $p=20$, $m=2$, $\tau=0.5$,
$\vect{b}=\vect{0}_{m}$ and $A=[C, O_{m \times (p-m)}]$ with
\begin{equation*}
	C=\left ( 
	\begin{array} {ccc}  
		3 & 0 \\    
		0 & 1
	\end{array}
	\right ).
\end{equation*}
The regression matrix $X$ has entries sampled from a standard normal
%Gaussian
random distribution, the null hypothesis is $H_0:
\ \vect{\beta}=\vect{0}$; we consider two alternative hypotheses
$$
H_{1,1}: \ \vect{\beta}=\delta \vect{e}_1 \quad {\rm and}\quad H_{1,2}: \ \vect{\beta}=\delta \vect{e}_2,
$$ where $\vect{e}_i$ is the $i$th coordinate (canonical) vector. With
the unrescaled $\infty$-S test, one expects the power to be low under
$H_{1,1}$ and good under $H_{1,2}$, while with the rescaled $\infty$-S
test, one expects the power to be high under both $H_{1,1}$ and
$H_{1,2}$. Figure~\ref{fig:rescaleTF} illustrates that it is really so.
%the improvement thanks to homopower rescaling.
While this example represents certainly just one very special case, we
remark that what we observe here also occurs naturally, owing to the
relation between $A$ and $X$ that is not controlled by the user
(except possibly in ANOVA). Thus, it is not only the $\infty$-S test,
but many other tests that would profit from such a rescaling in terms
of power.
%So homopower rescaling should be employed.

%%%%%%%%%%%%%%%%%%%%%%%%%%%%%
\subsection{The total variation $\infty$-S$^\tau$test}\label{sct:powerTV}

Total variation \citep{ROF92} aims at detecting jumps in a noisy
signal or a time series ${\bf y}=\vect{\beta}+\vect{e}$, which
corresponds to model~\eqref{eq:linearmodel} with $X=I_n$. A jump
occurs if $\sum_{j=1}^{n-1} |\beta_{j+1}-\beta_j|\neq 0$. Let $A$ be the
$(n-1)\times n$ (sparse) matrix with $-1$ and $1$ entries such that
$$
\sum_{j=1}^{n-1} |\beta_{j+1}-\beta_j|\ = \|A \vect{\beta} \|_1.
$$
Note that $A$ is of full row rank. To test for jump in the median or
any $\tau$-quantile, one can apply the $\infty$-S$^\tau$ test or the
$\infty$-rankS$^\tau$ test at the desired $\tau$-quantile.

\begin{figure}[!b]
	\centering
	\includegraphics[width=4.7in]{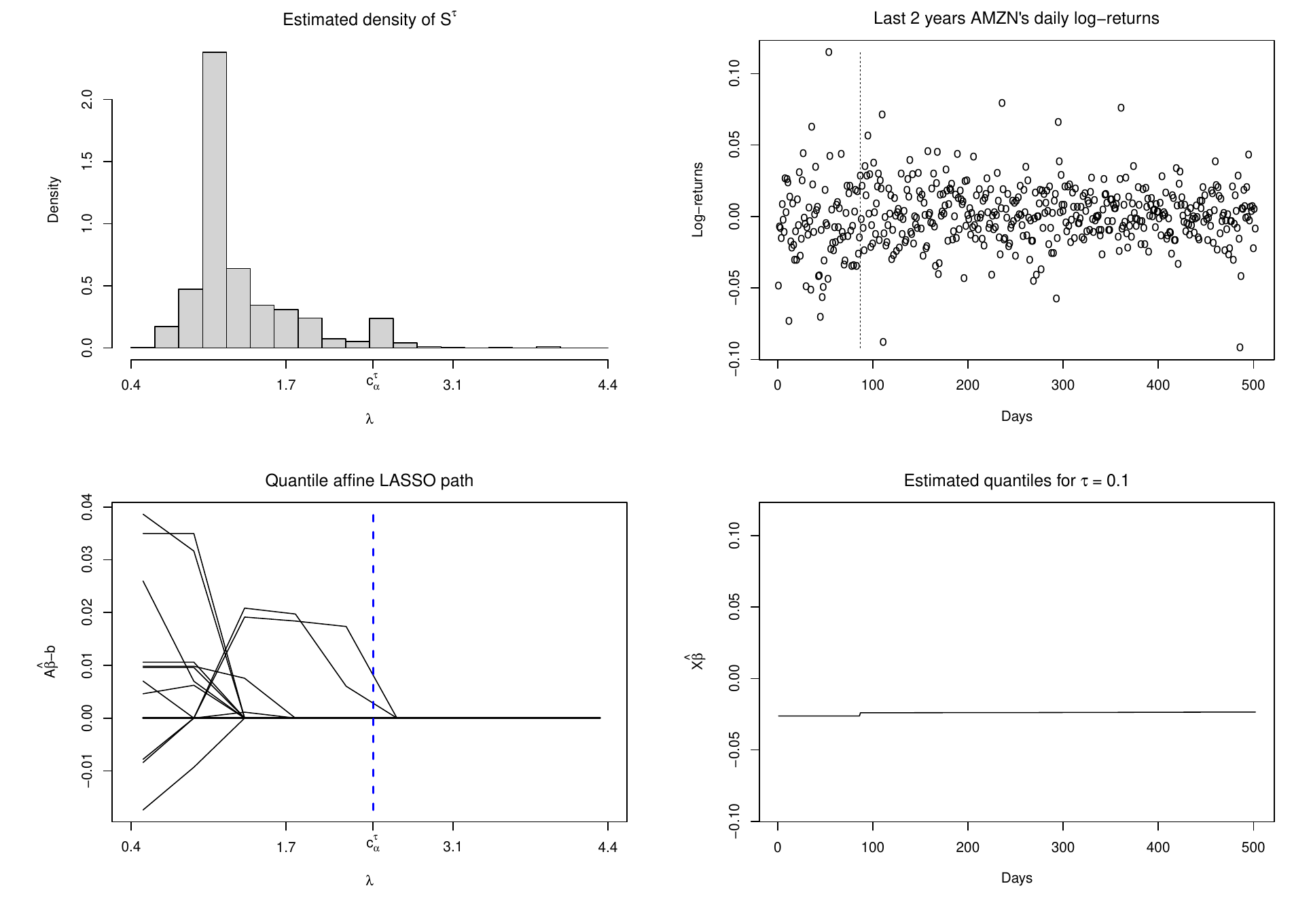}
	\caption{AMZN daily log-returm time series analysis. Top left plot: estimated density function of $f_{S^\tau}$ and critical value $c_\alpha^\tau$ for $\alpha=0.05$ by Monte Carlo simulation. Bottom left: regression quantile affine LASSO  path for $\tau=0.1$ with critical value $c_\alpha^\tau$.
% and zero-thresholding function $\lambda_0(\vect{y}, X, A)$ evaluated at~$\vect{y}$. 
Top right plot: AMZN time series data and estimate jump locations. Bottom right: estimated $\tau$-quantiles.}
	\label{fig:section53b}
\end{figure}

We apply the quantile total variation $\infty$-S$^\tau$ test to
analyze the time series of the Amazon daily log-returns. Focusing on
the lower tail which corresponds to drops in the AMZN stock value, we
test for constant $\tau$-quantile for $\tau=0.1$. The
$\infty$-S$^\tau$ test is rejected with an estimated p-value of
$0.0132$. The top left plot of Figure~\ref{fig:section53b} shows the
histogram estimate of the density $f_{S^\tau}$ of the test statistic
(by Monte Carlo sampling of the test statistic $S^\tau$ under the null
hypothesis) and the critical value $c_\alpha^\tau$ for $\alpha=0.05$.
Right below this plot, Figure~\ref{fig:section53b} shows the quantile
affine LASSO path, along with the critical value $c_\alpha^\tau$ 
%and the zero-thresholding function $\lambda_0^\tau(\vect{y},X,A)$
evaluated at the data $\vect{y}$. The right plots of
Figure~\ref{fig:section53b} show the raw AMZN time series data (top)
and the quantile affine LASSO estimate for $\lambda=c_\alpha^\tau$
(bottom), which corresponds to the quantile universal threshold
estimate \citep{Giacoetal17} of the $\tau$-quantile. The single jump
points to a potential change of regime moving to a less volatile
return.

\subsection{Robustness of the level to non-Gaussian errors}

Under the null $H_0: A{\vect \beta}={\vect b}$ and for a nominal level $\alpha=0.01$, we show the effective level of the $F$-test and the $\infty$-S test as a function of the degrees of freedom for Student errors d.f. $\in\{1,2,3,4,5,10,100\}$. We choose $n=100$, $p=20$,  $A$ is the first $m=5$ rows of the finite difference matrix of Section~\ref{sct:powerTV} and 
${\vect \beta}$ 
%=(3,3,4,4, \vect{h})$, where  $\vect{h}$
are $p$ samples from the standard Gaussian distribution; $\vect{b}$ is calculated as $A{\vect \beta}$.  %${\rm N}(0,5^2)$. %; so $A{\vect \beta}$ has two null entries.
In a robustness analysis of the F-test, \citet{ALI96} observe that major determinant of the sensitivity to nonnormality of the errors is the extent of the nonnormality of the regressors or the extent of presence of `leveraged' (influential) observations. So we simulate the entries of the matrix $X$ as a sample from $T$, where $T$ follows a student distribution with 2 degrees of freedom. 

\begin{figure}[!t]
	\centering
	\includegraphics[width=4.0in]{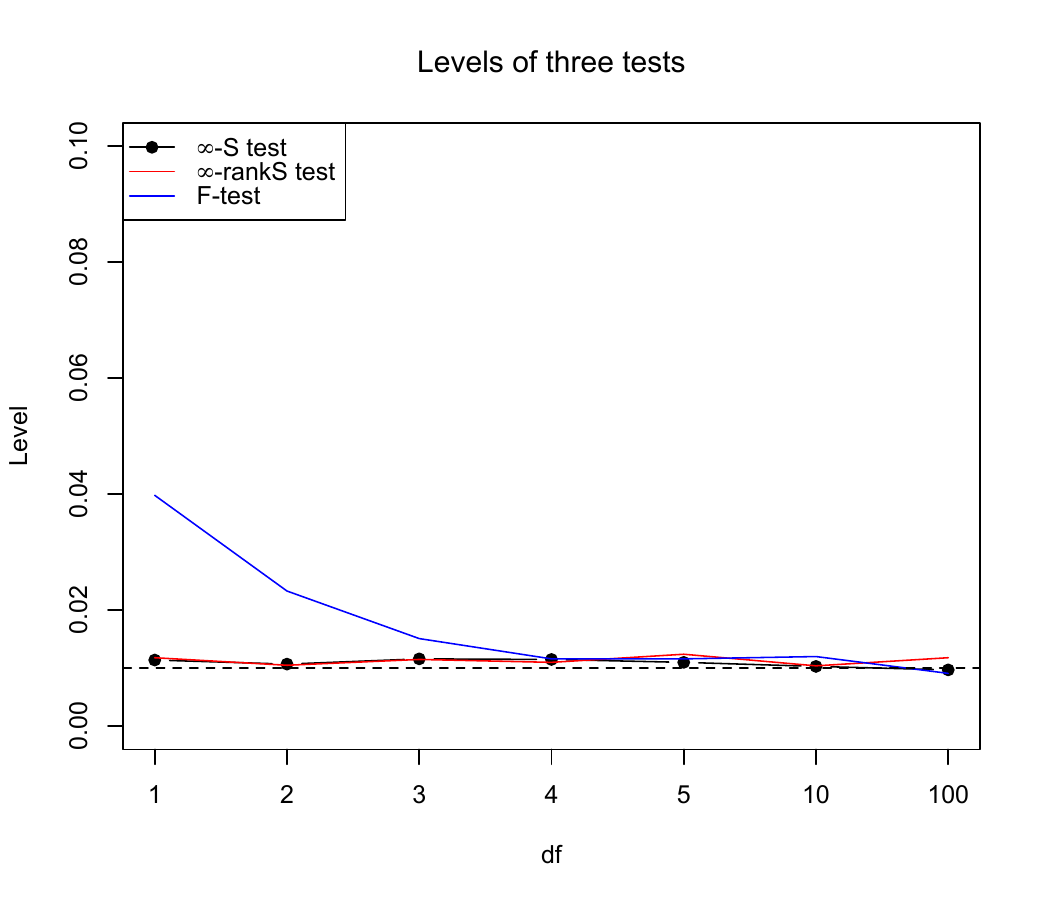}
	\caption{Level of the $\infty$-S test, $\infty$-rankS test and F-test as a function of the degrees of freedom of the Student errors. The dotted line is the nominal level $\alpha=0.01$.}
	\label{fig:binomialdensities}
\end{figure}

Over-shooting the nominal level leads to too many type I errors, hence rejecting too many  null hypotheses, that is, making too many false discoveries.
Here the two $\infty$-S tests match the nominal level, regardless of the degree of freedom of the Student errors. For this simulation, the  implementation of the {\tt quantreg} package does not return a p-value for $\chi^2$ sign test.

\section{Conclusions}

The $\infty$-S$^\tau$ test allows to test general linear null hypotheses for linear models and for any quantile $\tau$. The test function is simple to implement and can handle high-dimensional problems. We emphasize the $\infty$-S test which corresponds to the $\ell_1$-LASSO penalty, and showed that other norms than the $\ell_1$-norm could be employed, leading to other tests.
\citet{SARDY2022107507} show that using $\ell_1$ leads to high power under sparse alternatives (that is, most entries, but not all,  of $A\vect{\beta}-\vect{b}$ are zero when $H_1$ is true), and, under dense alternatives (that is, most  or all entries of $A\vect{\beta}-\vect{b}$ are non-zero), that the sign test based on $\ell_2$  has more power than the $\infty$-S test. To get nearly the best power with a single test regardless whether the alternative hypothesis is sparse or dense, \citet{SARDY2022107507} propose the $\oplus$-test.

The inversion of $\infty$-S test can be used to derive confidence
regions. The $(1-\alpha)$-confidence region for $A\vect{\beta}$ based
on the $\infty$-S test consists of all vectors~$\vect{b}$ that are not
rejected for $H_0 : A\vect{\beta} = \vect{b}$. In particular, to test
$H_0:\ \beta_j=b_j$ for some fixed $j\in\{1,\ldots,p \}$, one uses
$A=\vect{e}_j^{\rm T}$, the $j$th coordinate (canonical) vector; the
statistic is $s=|\vect{x}_j^{\rm T} \vect{\omega}|$, where
$\vect{\omega}={\rm sign}(\vect{y}-\vect{x}_jb_j - X_{-j} \hat
{\vect{\beta}}_{-j})$. So the $(1-\alpha)$-confidence interval for
$\beta_j$ is the set of all $b_j$ such that $|\vect{x}_j^{\rm T} {\rm
  sign}(\vect{y}-\vect{x}_jb_j - X_{-j} \hat {\vect{\beta}}_{-j}(b_j))
|\leq c_\alpha$, where $\hat{\vect{\beta}}_{-j}(b_j) \in
\arg\min_{\tilde{ \vect{\beta}} \in\mathbb{R}^{p-1}} \|
\vect{y}-\vect{x}_j b_j - X_{-j} \tilde{{\vect{\beta}}} \|_1$ and
$X_{-j}$ is the $X$ data matrix without its $j$th column.
%, the scalar product between the $j$th covariates and the sign of the residuals of the fit under $H_0$.

The $\infty$-S test is available and the research is made reproducible
in the {\tt Stest} package in {\tt R}, which can be downloaded from
\url{https://github.com/StatisticsL/Stest}.

\backmatter

%\bmhead{Supplementary information}
%
%We display the additional technical details in the
%supplementary material.

\bmhead{Acknowledgements}

The research of Sylvain Sardy was partially supported by the Swiss
National Science Foundation grant 200021E\_213166, a partner in the
Lead Agency project GA\v{C}R No.~23--06461K of the Czech Science
Foundation, which supported the research of Ivan Mizera; the research
of Xiaoyu Ma was supported by the National Natural Science Foundation of China(No. 12401397).%
%\section*{Declarations}
%
%Some journals require declarations to be submitted in a standardised format. Please check the Instructions for Authors of the journal to which you are submitting to see if you need to complete this section. If yes, your manuscript must contain the following sections under the heading `Declarations':
%
%\begin{itemize}
%\item Conflict of interest/Competing interests (check journal-specific guidelines for which heading to use)
%\item Ethics approval and consent to participate
%\item Consent for publication
%\item Data availability 
%\end{itemize}
%
%\noindent
%If any of the sections are not relevant to your manuscript, please include the heading and write `Not applicable' for that section. 
%
%%%===================================================%%
%%% For presentation purpose, we have included        %%
%%% \bigskip command. Please ignore this.             %%
%%%===================================================%%
%\bigskip
%\bigskip\noindent
%Springer journals and proceedings: \url{https://www.springer.com/gp/editorial-policies}
%
%\bigskip\noindent
%Nature Portfolio journals: \url{https://www.nature.com/nature-research/editorial-policies}
%
%\bigskip\noindent
%\textit{Scientific Reports}: \url{https://www.nature.com/srep/journal-policies/editorial-policies}
%
%\bigskip\noindent
%BMC journals: \url{https://www.biomedcentral.com/getpublished/editorial-policies}
%\end{flushleft}
%
%
%

\begin{appendices}

\section{Proof of Theorem~1}

Consider the primal cost function in~(2.5). This primal cost being convex,  the gap between the primal and dual cost functions is zero at optimality. The dual cost can be derived as follows
\begin{eqnarray*}
	&&\min_{\vect{\beta}\in{\mathbb R}^{p}}    \| \vect{y}- X\vect{\beta}\|_1 + \lambda \| A\vect{\beta}-\vect{b}\|_1\\
	&=& \min_{\vect{\beta}\in{\mathbb R}^{p}, \vect{u}\in{\mathbb R}^{n}}     \| {\vect u}\|_1 + \lambda \| A\vect{\beta}-\vect{b}\|_1 \quad {\rm s.t.} \quad  \vect{u}=\vect{y}- X\vect{\beta} \\
	&=& \min_{\vect{\beta}\in{\mathbb R}^{p}, \vect{u}\in{\mathbb R}^{n}}  \max_{{\vect \omega}}  \| {\vect u}\|_1 + \lambda \| A\vect{\beta}-\vect{b}\|_1  +  {\vect \omega}^{\rm T} (\vect{u}-\vect{y} + X\vect{\beta}) \\
	&=&  \max_{{\vect \omega}} -{\vect \omega}^{\rm T} \vect{y}  + \min_{ \vect{u}\in{\mathbb R}^{n}}  \| {\vect u}\|_1 + {\vect \omega}^{\rm T} \vect{u} + \min_{\vect{\beta}\in{\mathbb R}^{p}}  \lambda \| A\vect{\beta}-\vect{b}\|_1  +  {\vect \omega}^{\rm T}  X\vect{\beta} \\
	&=&  \max_{\| {\vect \omega} \|_\infty \leq 1, X^{\rm T}{\vect \omega}\perp {\rm ker}(A)} -{\vect \omega}^{\rm T} \vect{y}  + 0  +  \lambda \| A\vect{\beta}^\star-\vect{b}\|_1  +  {\vect \omega}^{\rm T}  X\vect{\beta}^\star, \\
	%  &=&  \max_{\| {\vect \omega} \|_\infty \leq 1} -{\vect \omega}^{\rm T} \vect{y}  +  \lambda \| A\vect{\beta}^\star-\vect{b}\|_1  +  {\vect \omega}^{\rm T}  X\vect{\beta}^\star,
\end{eqnarray*}
where $\vect{\beta}^\star$ satisfies the KKT conditions
\begin{equation}\label{eq:KKT}
	\lambda A^{\rm T} \gamma(A\vect{\beta}^\star-\vect{b})\ni X^{\rm T}  {\vect \omega},
\end{equation}
where $\gamma()$ applied componentwise to the penalized vector ${\vect \kappa}=A\vect{\beta}^\star-\vect{b}$ is defined by $\gamma(\kappa_k)={\rm sign}(\kappa_k)$ if $\kappa_k\neq 0$, and $\gamma(0)\in[-1,1]$, for $k=1,\ldots,m$.
We are interested in finding the smallest $\lambda$ for the solution to~(2.5) to satisfy $H_0$. In that case, ${\vect \kappa}={\vect 0}$ and $ {\vect{\beta}^\star=\hat{\vect{\beta}}}_{H_0}$, so $\gamma(A\vect{\beta}^\star-\vect{b})={\cal B}^m_{\infty}(1)$, the unit ball of the infinite norm in ${\mathbb R}^m$, and~\eqref{eq:KKT} is
%Under $H_0$, using the KKT conditions, the dual cost is 
%%$g({\vect \omega}^\star)=- \vect{y}^{\rm T}{\vect \omega}^\star+0+\lambda {\vect{\beta}^\star}^{\rm T}A^{\rm T}{\cal B}_{\infty}(1)=- \vect{y}^{\rm T}{\vect \omega}^\star+\lambda{\rm b}^{\rm T}{\cal B}_{\infty}(1)$
%$g({\vect \omega}^\star):=- {{\vect \omega}^\star}^{\rm T} \vect{y}+0+ {{\vect \omega}^\star}^{\rm T}  X\vect{\beta}^\star$, and the primal cost
%is $f(\vect{\beta}^\star):=\|\vect{y}- X\vect{\beta}^\star \|_1$ with $\vect{\beta}^\star=\hat{\vect{\beta}}_{H_0}$. Since the duality gap is zero, then $g({\vect \omega}^\star)=f(\vect{\beta}^\star)$, that is, $- {{\vect \omega}^\star}^{\rm T}( \vect{y} - X \hat{\vect{\beta}}_{H_0})=\|\vect{y}- X\hat{\vect{\beta}}_{H_0} \|_1$. Consequently ${\vect \omega}^\star=\gamma(X\hat{\vect{\beta}}_{H_0}-{\vect y})$, with the same definition of $\gamma()$ as above. The KKT conditions therefore are
\begin{equation}\label{eq:lambdaconditions}
	\lambda A^{\rm T}  {\cal B}^m_{\infty}(1)\ni X^{\rm T}{\vect \omega}.
\end{equation}
But $X^{\rm T}{\vect \omega}\perp {\rm ker}(A)$ and $A$ is full row rank, so there exists a unique ${\vect \alpha}\in {\mathbb R}^m$ such that $X^{\rm T}{\vect \omega}=A^{\rm T}{\vect \alpha}$; moreover ${\vect \alpha}=(AA^{\rm T})^{-1} A X^{\rm T}{\vect \omega}$ is this unique solution.
The smallest $\lambda$ satisfying conditions~\eqref{eq:lambdaconditions} is $\| {\vect \alpha}\|_\infty$, so
%$$
%\lambda   {\cal B}^m_{\infty}(1)\ni (AA^{\rm T})^{-1} X^{\rm T}{\vect \omega},
%$$
%where the inverse exists since $A$ is full row rank. So
%Consider the right hand-side of~\eqref{eq:lambdaconditions}.
%Let ${\vect r}=X{\hat{\vect{\beta}}_{H_0}}-{\vect y}$, let ${\cal I}_0=\{ i\in\{1,\ldots,n\}: r_i=0\}$
%and let $\vect{x}_{{\cal I}_0}\in{\mathbb R}^{p}$ be the $\ell_1$ norms of the $p$ rows of $(X^{\rm T})_{{\cal I}_0}$.
%Letting ${\vect u}:=(X^{\rm T})_{{\cal I}_0^{\rm c}}{\rm sign}({\vect r}_{{\cal I}_0^{\rm c}})$,
%since  $X^{\rm T}\gamma(X{\hat{\vect{\beta}}_{H_0}}-{\vect y})={\vect u}+(X^{\rm T})_{{\cal I}_0}{\cal B}^{|{\cal I}_0|}_{\infty}(1)$, the absolute value of the smallest components in magnitude of the right hand-side  are therefore the vector ${\vect v}:= \max(|{\vect u}|-\vect{x}_{{\cal I}_0},0)$, where all functions are applied componentwise.
%Letting $\vect{a}\in{\mathbb R}^p$ be the $\ell_1$ norms of the $p$ columns of $A$ and ${\cal J}_A=\{ j\in\{1,\ldots,p\}: a_j \neq 0\}$, we therefore have
$$
\lambda_0(\vect{y}, X, A)= \|(AA^{\rm T})^{-1} A X^{\rm T}{\vect \omega} \|_\infty.
$$
Under $H_0$, % using the KKT conditions, 
the dual cost is 
%$g({\vect \omega}^\star)=- \vect{y}^{\rm T}{\vect \omega}^\star+0+\lambda {\vect{\beta}^\star}^{\rm T}A^{\rm T}{\cal B}_{\infty}(1)=- \vect{y}^{\rm T}{\vect \omega}^\star+\lambda{\rm b}^{\rm T}{\cal B}_{\infty}(1)$
$g({\vect \omega}^\star):=- {{\vect \omega}^\star}^{\rm T} \vect{y}+0+ {{\vect \omega}^\star}^{\rm T}  X\vect{\beta}^\star$, and the primal cost
is $f(\vect{\beta}^\star):=\|\vect{y}- X\vect{\beta}^\star \|_1$ with $\vect{\beta}^\star=\hat{\vect{\beta}}_{H_0}$. Since the duality gap is zero, then $g({\vect \omega}^\star)=f(\vect{\beta}^\star)$, that is, $- {{\vect \omega}^\star}^{\rm T}( \vect{y} - X \hat{\vect{\beta}}_{H_0})=\|\vect{y}- X\hat{\vect{\beta}}_{H_0} \|_1$. Consequently ${\vect \omega}^\star=\gamma(X\hat{\vect{\beta}}_{H_0}-{\vect y})$, with the same definition of $\gamma()$ as above. Some entries ${{\cal I}\in\{1,\ldots,n\}}$ of the residuals ${\vect r}={\vect y}-X\hat{\vect{\beta}}_{H_0}$ are different from zero, in which case ${\vect \omega}^\star_{\cal I}={\rm sign}(-{\vect r}_I)$. Let now $K\in{\mathbb R}^{p\times (p-m)}$ be a basis for the kernel of $A$. We have that $K^{\rm T} X^{\rm T} {\vect \omega}={\vect 0}_{p-m}$.
Letting $X_{\cal I}$ be the rows of $X$ which indexes are in~${\cal I}$, $K^{\rm T} X^{\rm T} {\vect \omega}^\star=K^{\rm T} (X_{\cal I}^{\rm T} {\vect \omega}^\star_{\cal I}+ X_{{\cal I}^c}^{\rm T} {\vect \omega}^\star_{{\cal I}^c})\equiv 0$ iff $K^{\rm T} X_{{\cal I}^c}^{\rm T} {\vect \omega}^\star_{{\cal I}^c} = -K^{\rm T} X_{\cal I}^{\rm T} {\vect \omega}^\star_{\cal I}$. Solving this linear system leads to the remaining vector of the dual  ${\vect \omega}^\star_{{\cal I}^c} \in {\mathbb R}^{p-m}$.

\section{Proof of Theorem~2}

The residuals of the least absolute fit have a distribution which scale is proportional to $\xi$; so since $S= \|(AA^{\rm T})^{-1} A X^{\rm T}{\vect \omega} \|_\infty$ is a function of the dual vector which is the sign of the residuals, it is pivotal with respect to $\xi$. 
Moreover assuming the LAD estimate ${\hat{\vect{\beta}}_{H_0}}$ is asymptotically Gaussian with mean ${\vect \beta}_{H_0}$, we have that $X{\hat{\vect{\beta}}_{H_0}}-{\vect Y}=X(\hat{\vect{\beta}}_{H_0}-{\vect{\beta}_{H_0}})-X {\vect \epsilon}$ is asymptotically pivotal with respect to ${\vect{\beta}}_{H_0}$.

%%%%%%%%%%%%%%%%%%%%%%%%%

\section{Proof of Theorem~4}

It is well known that the LAD optimization problem (here with $\vect{y}=\vect{y}_{-A}$ and $X=X_{-A}$)  can be rewritten as the linear program:
$$
\min_{\vect{r}_+\geq 0, \vect{r}_-\geq 0,\vect{\beta} } (\vect{r}_+^{\rm T}, \vect{r}_-^{\rm T})^{\rm T} \vect{1}_{2n}\quad {\rm s.t.}\quad  \vect{y}_{-A}=X_{-A}\vect{\beta}_{-A}+\vect{r}_+-\vect{r}_-,%\ \vect{r}_+>0, \ \vect{r}_->0,
$$
where $\vect{1}_{2n}$ is the vector of ones of length $2n$. Using the Lagrange multiplier dual variable $\vect{\omega}$, the dual can be derived as follows.
\begin{eqnarray*}
	&&\min_{\vect{r}_+\geq 0, \vect{r}_-\geq 0,\vect{\beta}_{-A}} \max_{\vect{\omega}}(\vect{r}_+^{\rm T}, \vect{r}_-^{\rm T})^{\rm T} \vect{1}_{2n} + \vect{\omega}^{\rm T} ( \vect{y}_{-A}-X_{-A}\vect{\beta}_{-A}-\vect{r}_+ +\vect{r}_-)\\
	&=& \max_{\vect{\omega}}  \vect{\omega}^{\rm T}  \vect{y}_{-A} - \min_{\vect{\beta}_{-A}} \vect{\beta}_{-A}^{\rm T} X_{-A}^{\rm T}\vect{\omega} + \min_{\vect{r}_+\geq 0} (\vect{1}_n^{\rm T} \vect{r}_+ - \vect{\omega}^{\rm T} \vect{r}_+ )   + \min_{\vect{r}_-\geq0} (\vect{1}_n^{\rm T} \vect{r}_- + \vect{\omega}^{\rm T} \vect{r}_-) \\
	&=&\max_{\|\vect{\omega}\|_\infty\leq 1, X_{-A}^{\rm T}\vect{\omega}={\bf 0}}   \vect{\omega}^{\rm T}  \vect{y}_{-A} + 0
\end{eqnarray*}
with $w_i=1=$ when $(r_+)_i>0$, $w_i=-1$ when $(r_-)_i>0$, so $\vect{\omega}$ is the sign of the residuals $\vect{y}_{-A}-X_{-A}\hat{{\vect{\beta}}}_{-A}=\vect{y }-X \hat{\vect{\beta}}_{H_0}$ when the residuals are non-zero. So the dual problems of the constrained LAD and affine LAD-LASSO when $\lambda=\lambda_0(\vect{Y},X,A)$ are the same.

\section{Proof of Theorem~5}

The proof follows the same lines as the proof of Theorem~1. The dual cost to~(4.12) is the same except that the constraint on ${\vect \omega} \in [-\tau, 1-\tau]^n$ with the same KKT conditions.
We are interested in finding the smallest $\lambda$ for the solution to~(4.12) to satisfy $H_0$. In that case, ${\vect \kappa}={\vect 0}$ and $ {\vect{\beta}^\star=\hat{\vect{\beta}}}_{H_0}^\tau$, so $\gamma(A\vect{\beta}^\star-\vect{b})={\cal B}^m_{\infty}(1)$, the unit ball of the infinite norm in ${\mathbb R}^m$, and the KKT conditions are %~\eqref{eq:KKTrho} is
%Under $H_0$, using the KKT conditions, the dual cost is 
%%$g({\vect \omega}^\star)=- \vect{y}^{\rm T}{\vect \omega}^\star+0+\lambda {\vect{\beta}^\star}^{\rm T}A^{\rm T}{\cal B}_{\infty}(1)=- \vect{y}^{\rm T}{\vect \omega}^\star+\lambda{\rm b}^{\rm T}{\cal B}_{\infty}(1)$
%$g({\vect \omega}^\star):=- {{\vect \omega}^\star}^{\rm T} \vect{y}+0+ {{\vect \omega}^\star}^{\rm T}  X\vect{\beta}^\star$, and the primal cost
%is $f(\vect{\beta}^\star):=\|\vect{y}- X\vect{\beta}^\star \|_1$ with $\vect{\beta}^\star=\hat{\vect{\beta}}_{H_0}$. Since the duality gap is zero, then $g({\vect \omega}^\star)=f(\vect{\beta}^\star)$, that is, $- {{\vect \omega}^\star}^{\rm T}( \vect{y} - X \hat{\vect{\beta}}_{H_0})=\|\vect{y}- X\hat{\vect{\beta}}_{H_0} \|_1$. Consequently ${\vect \omega}^\star=\gamma(X\hat{\vect{\beta}}_{H_0}-{\vect y})$, with the same definition of $\gamma()$ as above. The KKT conditions therefore are
\begin{equation}\label{eq:lambdaconditions1}
	\lambda A^{\rm T}  {\cal B}^m_{\infty}(1)\ni X^{\rm T}{\vect \omega}.
\end{equation}
But $X^{\rm T}{\vect \omega}\perp {\rm ker}(A)$ and $A$ is full row rank, so there exists a unique ${\vect \alpha}\in {\mathbb R}^m$ such that $X^{\rm T}{\vect \omega}=A^{\rm T}{\vect \alpha}$; moreover ${\vect \alpha}=(AA^{\rm T})^{-1} A X^{\rm T}{\vect \omega}$ is this unique solution.
The smallest $\lambda$ satisfying conditions~\eqref{eq:lambdaconditions1} is $\| {\vect \alpha}\|_\infty$, so
%$$
%\lambda   {\cal B}^m_{\infty}(1)\ni (AA^{\rm T})^{-1} X^{\rm T}{\vect \omega},
%$$
%where the inverse exists since $A$ is full row rank. So
%Consider the right hand-side of~\eqref{eq:lambdaconditions}.
%Let ${\vect r}=X{\hat{\vect{\beta}}_{H_0}}-{\vect y}$, let ${\cal I}_0=\{ i\in\{1,\ldots,n\}: r_i=0\}$
%and let $\vect{x}_{{\cal I}_0}\in{\mathbb R}^{p}$ be the $\ell_1$ norms of the $p$ rows of $(X^{\rm T})_{{\cal I}_0}$.
%Letting ${\vect u}:=(X^{\rm T})_{{\cal I}_0^{\rm c}}{\rm sign}({\vect r}_{{\cal I}_0^{\rm c}})$,
%since  $X^{\rm T}\gamma(X{\hat{\vect{\beta}}_{H_0}}-{\vect y})={\vect u}+(X^{\rm T})_{{\cal I}_0}{\cal B}^{|{\cal I}_0|}_{\infty}(1)$, the absolute value of the smallest components in magnitude of the right hand-side  are therefore the vector ${\vect v}:= \max(|{\vect u}|-\vect{x}_{{\cal I}_0},0)$, where all functions are applied componentwise.
%Letting $\vect{a}\in{\mathbb R}^p$ be the $\ell_1$ norms of the $p$ columns of $A$ and ${\cal J}_A=\{ j\in\{1,\ldots,p\}: a_j \neq 0\}$, we therefore have
$$
\lambda_0^\tau(\vect{y}, X, A)= \|(AA^{\rm T})^{-1} A X^{\rm T}{\vect \omega} \|_\infty.
$$
Under $H_0$, % using the KKT conditions, 
the dual cost is 
%$g({\vect \omega}^\star)=- \vect{y}^{\rm T}{\vect \omega}^\star+0+\lambda {\vect{\beta}^\star}^{\rm T}A^{\rm T}{\cal B}_{\infty}(1)=- \vect{y}^{\rm T}{\vect \omega}^\star+\lambda{\rm b}^{\rm T}{\cal B}_{\infty}(1)$
$g({\vect \omega}^\star):=- {{\vect \omega}^\star}^{\rm T} \vect{y}+0+ {{\vect \omega}^\star}^{\rm T}  X\vect{\beta}^\star$, and the primal cost
is $f(\vect{\beta}^\star):=\| \vect{y}- X\vect{\beta}^\star\|_{\rho_\tau}$ with $\vect{\beta}^\star=\hat{\vect{\beta}}_{H_0}^\tau$. Since the duality gap is zero, then $g({\vect \omega}^\star)=f(\vect{\beta}^\star)$, that is, $- {{\vect \omega}^\star}^{\rm T}{\vect r} =\|\vect{r} \|_{\rho_\tau}$ with the residuals ${\vect r}=\vect{y} - X \hat{\vect{\beta}}_{H_0}^\tau $. Consequently ${\vect \omega}^\star=\gamma_\tau({\bf r})$ with  $\gamma_\tau(r_k)=1-\tau$ if $r_k< 0$, $\gamma_\tau(r_k)=-\tau$ if $r_k> 0$ and $\gamma_\tau(0)\in[\tau-1,\tau]$, for $k=1,\ldots,m$.
Some entries ${{\cal I}\in\{1,\ldots,n\}}$ of the residuals ${\vect r}$ are different from zero, in which case ${\vect \omega}^\star_{\cal I}=\gamma_\tau({\vect r}_I)$. Let now $K\in{\mathbb R}^{p\times (p-m)}$ be a basis for the kernel of $A$. We have that $K^{\rm T} X^{\rm T} {\vect \omega}={\vect 0}_{p-m}$.
Letting $X_{\cal I}$ be the rows of $X$ which indexes are in~${\cal I}$, $K^{\rm T} X^{\rm T} {\vect \omega}^\star=K^{\rm T} (X_{\cal I}^{\rm T} {\vect \omega}^\star_{\cal I}+ X_{{\cal I}^c}^{\rm T} {\vect \omega}^\star_{{\cal I}^c})\equiv 0$ iff $K^{\rm T} X_{{\cal I}^c}^{\rm T} {\vect \omega}^\star_{{\cal I}^c} = -K^{\rm T} X_{\cal I}^{\rm T} {\vect \omega}^\star_{\cal I}$. Solving this linear system leads to the remaining vector of the dual  ${\vect \omega}^\star_{{\cal I}^c} \in {\mathbb R}^{p-m}$.

\section{Proof of Theorem~6}

The residuals of the least $\rho_\tau$ fit have a distribution which scale is proportional to $\xi$; so since $S^\tau= \|(AA^{\rm T})^{-1} A X^{\rm T}{\vect \omega}^\tau \|_\infty$ is a function of the dual vector which is the sign of the residuals, it is pivotal with respect to $\xi$. 
Moreover assuming the quantile regression estimate ${\hat{\vect{\beta}}^\tau_{H_0}}$ is asymptotically Gaussian with mean ${\vect \beta}^\tau_{H_0}$, 
we have that $X{\hat{\vect{\beta}}_{H_0}^\tau}-{\vect Y}=X(\hat{\vect{\beta}}_{H_0}^\tau-{\vect{\beta}_{H_0}^\tau})-X {\vect \epsilon}$ is asymptotically pivotal with respect to ${\vect{\beta}}_{H_0}^\tau$.

\end{appendices}


%\bibliography{article_bis}% common bib file
\bibliography{allbib}% common bib file

@Manual{Rogerrq,
    title = {quantreg: Quantile Regression},
    author = {Koenker, R.},
    year = {2024},
    note = {R package version 5.98},
    url = {https://CRAN.R-project.org/package=quantreg},
  }

@article{Roger93,
author = {Gutenbrunner, C. and Jure\v{c}kov\'a, J. and Koenker, R. and Portnoy, S.},
title = {Tests of linear hypotheses based on regression rank scores},
journal = {Journal of Nonparametric Statistics},
volume = {2},
number = {4},
pages = {307--331},
year = {1993},
publisher = {Taylor \& Francis}
}

@article{CDS99,
author = "Chen, S.~S. and Donoho, D.~L. and Saunders, M.~A.",
year = "1999",
title = "Atomic decomposition by basis pursuit",
journal = "SIAM Journal on Scientific Computing",
volume = "20(1)",
pages = "33--61"
}

@article{PG89,
author = "Di Pillo, G. and Grippo, L.",
year = "1989",
title = "Exact penalty functions in constrained optimization",
journal = "SIAM Journal on Control and Optimization",
volume = "27(6)",
pages = "1333--1360"
}

@article{SARDY2022107507,
title = {Thresholding tests based on affine LASSO to achieve non-asymptotic nominal level and high power under sparse and dense alternatives in high dimension},
journal = {Computational Statistics \& Data Analysis},
volume = {173},
pages = {107507},
year = {2022},
author = {Sardy, S. and Diaz-Rodriguez, J. and Giacobino, C.},
}

@article{Giacoetal17,
	Author = {Giacobino, C. and Sardy, S. and Diaz Rodriguez, J. and Hengartner, N.},
	Journal = {Electronic Journal of Statistics},
	Title = {Quantile universal threshold},
	volume={11},
	number = {2},
	pages = {4701--4722},
	Year = {2017}}

@article{ALI96,
title = {Robustness to nonnormality of regression {F}-tests},
journal = {Journal of Econometrics},
volume = {71},
number = {1},
pages = {175-205},
year = {1996},
author = {Ali, M.~M. and Sharma, S.~C.},
}

@inproceedings{brown1951median,
  title={On median tests for linear hypotheses},
  author={Brown, G.~W. and Mood, A.~M.},
  booktitle={Proceedings of the Second Berkeley Symposium on Mathematical Statistics and Probability},
  volume={2},
  pages={159--167},
  year={1951},
  organization={University of California Press}
}

@article{KB78,
 author               = {Koenker, R. and Bassett, G.},
 journal              = {Econometrica},
 pages                = {33--50},
 title                = {Regression Quantiles},
 volume               = {46},
 number               = {1},
 year                 = {1978},
 }

@article{LADasym91,
 author = {Pollard, D.},
 journal = {Econometric Theory},
 number = {2},
 pages = {186--199},
 publisher = {Cambridge University Press},
 title = {Asymptotics for Least Absolute Deviation Regression Estimators},
 volume = {7},
 year = {1991}
}

@article{ROF92,
 author               = {Rudin, L.~I. and Osher, S. and Fatemi, E.},
 journal              = {Physica D},
 pages                = {259-268},
 title                = {Nonlinear total variation based noise removal algorithms},
 volume               = {60},
 year                 = {1992},
 }

@article{SardyISI08,
 author               = {Sardy, S.},
 journal              = {International Statistical Review},
 pages                = {285--297},
 title                = {On the practice of rescaling covariates},
 volume               = {76},
 number               = {2},
 year                 = {2008},
 }

@article{Tibs:regr:1996,
 author               = {Tibshirani, R.},
 journal              = {Journal of the Royal Statistical Society, Series B},
 pages                = {267--288},
 title                = {Regression Shrinkage and Selection Via the Lasso},
 volume               = {58},
 number               = {1},
 year                 = {1996},
 }

@article{CIS-215377,
 author               = {Wang, H. and Li, G. and Jiang, G.},
 journal              = {Journal of Business \& Economic Statistics},
 number               = {3},
 pages                = {347--355},
 title                = {Robust Regression Shrinkage and Consistent Variable Selection Through the {LAD}-Lasso},
 volume               = {25},
 year                 = {2007},
 }

@article{Yuan:Lin:mode:2006,
 author               = {Yuan, M. and Lin, Y.},
 journal              = {Journal of the Royal Statistical Society, Series B},
 number               = {1},
 pages                = {49--67},
 title                = {Model Selection and Estimation in Regression with Grouped Variables},
 volume               = {68},
 year                 = {2006},
 }

@book{con99,
  title={Practical nonparametric statistics},
  author={Conover, William Jay},
  year={1999},
  publisher={John Wiley \& Sons},
  address={New York}
}

@book{spr89,
  author={Sprent, Peter},
  title={Introducing nonparametric methods},
  year={1989},
  publisher={Springer},
  address={New York}
}

@article{he2023,
  title={Smoothed quantile regression with large-scale inference},
  author={He, Xuming and Pan, Xiaoou and Tan, Kean Ming and Zhou, Wen-Xin},
  journal={Journal of Econometrics},
  volume={232},
  number={2},
  pages={367--388},
  year={2023},
  publisher={Elsevier}
}

@article{kolmog31,
  title={The method of the median in the theory of errors},
  author={Kolmogorov, Alexander Nikolaevich},
  journal={Matematicheski\u\i{} Sbornik},
  volume={38},
  pages={47--50},
  year={1931}
}

@book{koenker,
  title={Quantile regression},
  author={Koenker, Roger},
  year={2005},
  publisher={Cambridge University Press},
  address={Cambridge}
}

@article{gutejure92,
  title={Regression rank scores and regression quantiles},
  author={Gutenbrunner, Christoph and Jure\v{c}kov{\'a}, J},
  journal={The Annals of Statistics},
  volume={20},
  number={1},
  pages={305--330},
  year={1992},
  publisher={Institute of Mathematical Statistics}
}

\vspace{2cm} 
\textbf{Sylvian Sardy}, Section of Mathematics, University of Geneva,  rue du Conseil-Général 7-9, 1205 Geneva, Switzerland. 

\textbf{Ivan Mizera}, Department of Probability and Mathematical Statistics, Faculty of Mathematics and Physics, Charles University Prague, Sokolovsk\'a 83, 186 75 Praha, Czechia.

\textbf{Xiaoyu Ma}, College of Science, National University of Defense and Technology,  1 Fuyuan Road, 410000 Changsha, China. 

\textbf{Hugo Gaible}, Department of Mathematics, École Normale Supérieur Paris-Saclay, 4 avenue des Sciences,  91190 Gif-sur-Yvette, France. 

\end{document}